\documentclass[showpacs,epsfig,preprintnumbers,amsmath,amssymb,nofootinbib,20pt]{revtex4-1}
\usepackage{graphicx}
\usepackage{dcolumn}
\usepackage{bm}
\usepackage{subfigure}

\begin{document}
\def\d{{\rm d}}
\def\Epos{E_{\rm pos}}
\def\ap{\approx}
\def\eff{{\rm eft}}
\def\L{{\cal L}}
\newcommand{\vev}[1]{\langle {#1}\rangle}
\newcommand{\CL}   {C.L.}
\newcommand{\dof}  {d.o.f.}
\newcommand{\eVq}  {\text{EA}^2}
\newcommand{\Sol}  {\textsc{sol}}
\newcommand{\SlKm} {\textsc{sol+kam}}
\newcommand{\Atm}  {\textsc{atm}}
\newcommand{\Chooz}{\textsc{chooz}}
\newcommand{\Dms}  {\Delta m^2_\Sol}
\newcommand{\Dma}  {\Delta m^2_\Atm}
\newcommand{\Dcq}  {\Delta\chi^2}
\newcommand{\nbb}{$\beta\beta_{0\nu}$ }
\newcommand {\be}{\begin{equation}}
\newcommand {\ee}{\end{equation}}
\newcommand {\ba}{\begin{eqnarray}}
\newcommand {\ea}{\end{eqnarray}}
\def\VEV#1{\left\langle #1\right\rangle}
\let\vev\VEV
\def\e6{E(6)}
\def\10{SO(10)}
\def\21{SA(2) $\otimes$ U(1) }
\def\321{$\mathrm{SU(3) \otimes SU(2) \otimes U(1)}$ }
\def\lr{SA(2)$_L \otimes$ SA(2)$_R \otimes$ U(1)}
\def\422{SA(4) $\otimes$ SA(2) $\otimes$ SA(2)}
\newcommand{\AHEP}{%
School of physics, Institute for Research in Fundamental Sciences
(IPM)\\P.O.Box 19395-5531, Tehran, Iran\\

  }
\newcommand{\Tehran}{%
School of physics, Institute for Research in Fundamental Sciences (IPM)
\\
P.O.Box 19395-5531, Tehran, Iran}
\def\roughly#1{\mathrel{\raise.3ex\hbox{$#1$\kern-.75em
      \lower1ex\hbox{$\sim$}}}} \def\lsim{\roughly<}
\def\gsim{\roughly>}
\def\ltap{\raisebox{-.4ex}{\rlap{$\sim$}} \raisebox{.4ex}{$<$}}
\def\gtap{\raisebox{-.4ex}{\rlap{$\sim$}} \raisebox{.4ex}{$>$}}
\def\lsim{\raise0.3ex\hbox{$\;<$\kern-0.75em\raise-1.1ex\hbox{$\sim\;$}}}
\def\gsim{\raise0.3ex\hbox{$\;>$\kern-0.75em\raise-1.1ex\hbox{$\sim\;$}}}



\title{Secret  interactions of neutrinos with light gauge boson at the DUNE near detector }

\date{\today}
\author{P. Bakhti}\email{pouya\_bakhti@ipm.ir}
\author{Y. Farzan}\email{yasaman@theory.ipm.ac.ir}
\author{M. Rajaee}\email{meshkat.rajaee@ipm.ir}
\affiliation{\Tehran}
\begin{abstract}
	Secret interactions of neutrinos with light new gauge bosons, $Z^\prime$, can lead to a rich phenomenology in a supernova explosion as well as in the early Universe.  This interaction can also lead to new decay modes for charged mesons, $\pi^+ (K^+) \to e^+ \nu Z'$, and subsequently to $Z'\to \nu \bar{\nu}$. After demonstrating that such an interaction can be accommodated within viable electroweak symmetric models, we  study how the near detector (ND) of DUNE can probe this scenario. We also discuss how the DUNE ND can make it possible to reconstruct the flavor structure of the $Z^\prime$ coupling to neutrinos.
\end{abstract}
{}
\date{\today}
\maketitle
\section{Introduction}

The state-of-the-art DUNE experiment, which is mainly proposed to measure the CP-violating phase in the lepton sector, is a long baseline setup with two detectors.
The far detector (FD) will be located at a distance of 1300 km from the source at Sanford Underground Research Facility (SURF). The near detector (ND) will be at the site of Fermilab at a distance of $\sim$570 m from the source.  Having superb particle recognition and energy-momentum measurement precision  and receiving a high flux of neutrinos from the source make the near detector an ideal site to search for new light particles coupled to leptons. For example, Ref. \cite{Ballett:2018fah} shows that the ND of DUNE can be sensitive to MeV range sterile neutrinos that mix with $\nu_\mu$ and/or $\nu_e$. Refs. \cite{Coloma:2015pih} discusses the possibility of probing light dark matter particles produced at the source via a new gauge interaction involving baryons as well as dark matter particles.

In the present paper, we study the scenario in which neutrinos couple to a light gauge boson $Z'$ with a mass smaller than $\sim 100$ MeV. The new gauge boson can be produced via $\pi^+ \to l^+ \nu Z'$ or via $K^+ \to l^+ \nu Z'$ (where $l^+=e^+$ or $\mu^+$) and subsequently decay into a pair of a neutrino and an anti-neutrino before reaching the near detector. The produced neutrinos can be detected in the ND, providing us with information on the intermediate $Z'$.

Scenarios in which a light gauge boson couples to neutrinos are motivated within various contexts. For example, as shown in Ref. \cite{Aarssen:2012fx}, it can help to solve the missing satellite problem that is considered  a drawback for the canonical cold  dark matter scenarios. As demonstrated in Refs. \cite{Das:2017iuj,Dighe:2017sur}, such an interaction can lead to nontrivial features in the propagation of neutrinos within a supernova and, as a result, affect their energy spectrum. If such a coupling exists in nature but we overlook them in analyzing supernova or cosmological data, we will be lead to the wrong conclusion. It is therefore imperative to probe the possible new couplings of neutrinos by terrestrial experiments as much as we can.  If the $Z'$ also couples to matter fields, its effect can be probed by scattering experiments \cite{COHERENT} or by its impact on neutrino oscillation in matter. However, if the only standard particles that couple to $Z'$ are neutrinos, it will not have any impact on neutrino oscillation in matter or  on 
elastic scattering of neutrinos off nuclei \footnote{Notice that, even in the absence of coupling of $Z'$ to quarks, new scattering modes such as $\nu+q \to \nu+Z'+q$ or $\nu+q \to l+q'+Z'$ will open, but the rest of these inelastic scattering modes are suppressed by $g^2/16\pi^2$}.

From a model building point of view, it is straightforward to build a model giving rise to this scenario. As shown in Refs. \cite{Farzan:2016wym,Farzan:2017xzy}, if there is a sterile neutrino that is charged under a new $U(1)$ gauge symmetry  and is mixed with active neutrinos, the active neutrinos will also obtain a coupling to the gauge boson of the new $U(1)$ gauge symmetry. For completeness, we will outline the features of these models and discuss when the sterile neutrinos can be integrated out.

As shown in Ref.  \cite{Bakhti:2017jhm}, the meson decay experiments such as NA62 and PIENU can also search for such $Z'$ (see also \cite{Barger:2011mt,Belotsky:2001fb,Laha:2013xua,Ibe:2016dir}). In these experiments, the charged lepton produced in the decay of the charged meson can be identified, but the prompt neutrino, as well as the neutrino-antineutrino pair produced in the decay of $Z'$ goes undetected. As a result, these meson decay experiments are sensitive to $\sum_{\alpha \in \{ e,\mu,\tau ,[s]\}} |g_{e \alpha}|^2$ and $\sum_{\alpha \in \{ e,\mu,\tau ,[s]\}} |g_{\mu \alpha}|^2$ [in which $\nu_s$ is a possible  sterile neutrino which couples to $Z'$ along with $\nu_\mu$ or $\nu_e$ and is lighter than the decaying meson].  On the other hand, the charged lepton at the source of DUNE will not be detected, but, instead, the produced neutrino can be detected at the ND providing complementary information on the flavor structure of $Z'$ coupling.

The present paper is organized as follows. In Sec. \ref{model}, we describe two classes of  $U(1)$ gauge models that give rise to a neutrino interaction with $Z'$. 
In Sec. \ref{spectrum}, we compute the spectrum of neutrinos from $\pi^+ \to l^+ \nu Z'$ and subsequent $Z'\to \nu \bar{\nu}$ in the lab frame. In Sec. \ref{DUNEres}, constraints on the coupling of $Z^\prime$ with neutrinos are discussed using the DUNE near detector. Section \ref{Summary} is devoted to the summary and discussion.

\section{Models for neutrino interactions with new light gauge boson\label{model}}
In this section, we briefly review two classes of models that give rise to interactions of the type
\be \sum_{\alpha ,\beta} g_{\alpha \beta}Z'_\mu\bar{\nu}_\alpha \gamma^\mu \nu_\beta . \ee
As is well known, such an interaction can come from gauging various combination of lepton flavors and baryon number $a_e L_e+  a_\mu L_\mu+a_\tau L_\tau+b B$, where $a_e$, $a_\mu$, $a_\tau$, and $b$ are arbitrary real numbers that satisfy the anomaly cancellation condition $a_e+a_\mu+a_\tau+3b=0$. For such a gauge interaction, $g_{\alpha \alpha}\propto a_\alpha$ and $g_{\alpha \beta}|_{\alpha \ne \beta}=0$.
Famous combinations which have been extensively studied in the literature are $B-L=B-L_e-L_\mu-L_\tau$, $L_\mu -L_\tau$, $L_e -L_\tau$
and $L_\mu -L_e$. Notice that in these models, along with neutrinos, the corresponding charged leptons also couple to $Z'$. There are strong bounds on the coupling of the electron to $Z'$ from various
observations across a wide range of $Z'$ masses.
Because of the strong bounds on the electron coupling to $Z'$, we will not emphasize this class of models any further.

We now discuss another possibility for model building  which was proposed in Refs. \cite{Farzan:2016wym,Farzan:2017xzy}. Let us introduce a Dirac fermion $\Psi$  charged under the  new $U(1)$ and mixed with $\nu_\alpha$. Denoting the gauge coupling by $g_\Psi$, the gauge interaction term is $g_\Psi Z'_\mu \bar{\Psi} \gamma^\mu \Psi$.
 Since $\Psi$ mixes with active neutrinos, the active neutrinos of flavor $\nu_\alpha$ will be  a linear combination of mass eigenstates $\nu_i$: \be \label{nuA}\nu_\alpha =\sum_{i=1}^4 U_{\alpha i} \nu_i\ee in which $\nu_4$ is the heavier state that gives the main contribution to $\Psi$; {\it i.e.,} $U_{\Psi 4} \simeq 1$, $U_{\alpha 4}|_{\alpha=e,\mu,\tau} \ll 1$. Since $\nu_4$ is taken to be heavier than the charged meson $M^+$, in decay  $M^+ \to l_\alpha^+ \nu+X$ (where $X$ can be any state), the coherent $\nu$ state will not exactly be $\nu_\alpha$ [as defined in Eq. (\ref{nuA})] but will coincide with a linear  combination of $\nu_1$, $\nu_2$ and $\nu_3$ which cannot  be perpendicular to $\Psi$. In particular, for $M^+ \to l_\alpha^+ \nu$, it will be $\sum_{i=1}^3 U_{\alpha i} \nu_i$  so $\langle \Psi|\sum_{i=1}^3 U_{\alpha i}\nu_i \rangle=-U_{\Psi 4}^*U_{\alpha 4}$. Integrating out the heavy fourth state, the light active neutrinos receive a coupling of the form \footnote{In Ref. \cite{Farzan:2016wym}, $U_{\alpha 4}$ was denoted by $\kappa_\alpha$.}
$$ g_{\beta\alpha } Z_\mu' \bar{\nu}_\beta \gamma^\mu \nu_\alpha $$
in which
$$ g_{\beta\alpha }=g_\Psi|U_{\Psi 4}|^2 U_{\alpha 4}U_{\beta 4}^* \simeq  g_\Psi U_{\alpha 4}U_{\beta 4}^*. $$ As a result, three-body decays such as $\pi^+ ({\rm or}~K^+) \to l_{\alpha}^+ \nu_\beta Z'$ can take place with a rate proportional to $|g_{\alpha \beta}|^2$. $Z'$ will subsequently decay into $\bar{\nu}_\alpha \nu_\beta$ again with a rate proportional to  $|g_{\alpha \beta}|^2$.
However if $\nu_4$ is lighter than the parent charged lepton, we cannot integrate it out and the picture will be different. In this case, the neutrino state produced at the charged meson decay via standard model (SM) interactions will be
$\nu_\alpha =\sum_{i=1}^4 U_{\alpha i} \nu_i$ but if $\nu_4$ is heavy enough, it will decohere from $\nu_1$, $\nu_2$ and $\nu_3$, giving rise to $\Psi$ production given by $|U_{\alpha 4}|^2|U_{\psi 4}|^2\simeq |U_{\alpha 4}|^2$ but this is not the case we are interested in here. For our case with $m_\Psi \sim 1$ GeV,  $W^+ \to l^+_\alpha \nu_\alpha$ will take place with a rate deviated from the standard model prediction by a small amount of $O[|U_{\alpha 4}|^2(m_\Psi^2)/m_W^2]\ll 1$. The $\nu_4$ component in  $\nu_\alpha$ will subsequently decay into $\nu_i |_{i<4}$ and $Z'$ which appears again as missing energy at colliders.

As discussed in Ref. \cite{Farzan:2017xzy}, there are at least two mechanisms for mixing $\Psi$ with $\nu_\alpha$. In one method which is described in detail in Ref. \cite{Farzan:2016wym}, a new Higgs doublet $H^\prime$ charged under the new $U(1)$ is introduced such that its  vacuum expectation value induces a mixing between $\Psi$ and $\nu_\alpha$ via a Yukawa coupling of the form $\bar{L}_\alpha H^{\prime T}c\Psi$. In the other model described in detail in Sec. IV.B in Ref.  \cite{Farzan:2017xzy}, no new Higgs doublet is required. Instead, a neutral Dirac $N$ and a new scalar singlet $S$ charged under $U(1)$ are introduced with  interaction terms similar to that in the inverse seesaw mechanism: $Y_\alpha \bar{N}_RH^TcL_\alpha+\lambda_LS \bar{\Psi}_RN_L$. The mixing $U_{\alpha 4}$ will be given by $Y_\alpha \langle H \rangle \lambda_L \langle S\rangle/( m_Nm_\Psi)$.

Notice that in this class of models at tree level only neutrinos couple to $Z'$ so they are free from the bounds on the coupling of the corresponding charged leptons to $Z'$. In general, we expect $|g_{\alpha \beta}|=|g_{\beta \alpha}|$. Depending on the sign of $g_\Psi$, $g_{\alpha \alpha}$ can be either positive or negative. If there is only one $\Psi$ mixed with $\nu_\alpha$ and $\nu_\beta$, we expect $|g_{\alpha \beta}|^2= |g_{\alpha \alpha}|^2|g_{\beta \beta}|^2$ and if there is more than one $\Psi$, the Schwartz inequality implies $|g_{\alpha \beta}|^2< |g_{\alpha \alpha}|^2|g_{\beta \beta}|^2$. There are strong bounds on the deviation of the $3 \times 3$ PMNS mixing matrix from the unitarity that can be translated into the bounds on $g_{\alpha \beta}$
\cite{Farzan:2016wym}.  In this paper, we are mainly interested in the $g_{ee}$   and $g_{e\tau}$ elements because they can lead to kinematically favored decay modes $\pi^+ \to e^+ \nu_e Z'$ and $\pi^+ \to e^+ \nu_\tau Z'$,  respectively. To obtain nonzero $g_{ee}$, it is enough to mix $\Psi$ only with $\nu_e$. For gauge coupling in the perturbative range ({\it e.g.,} $g_\Psi\stackrel{<}{\sim}4$) the bound from unitarity ({\it i.e.,} $|U_{e4}|^2<2.5 \times 10^{-3}$ \cite{Fernandez-Martinez:2016lgt}) leads to $g_{ee} \stackrel{<}{\sim} 10^{-2}$. Notice that in this case, we are not introducing a new source of lepton flavor violating (LFV) so no strong bound comes from $\mu \to e \gamma$ and from similar LFV processes. The unitarity  bound on $|U_{e4}U_{\tau 4}^*|$ is $3.7 \times 10^{-3}$ \cite{Fernandez-Martinez:2016lgt} which  again translates into $g_{e\tau}\stackrel{<}{\sim}10^{-2}$.
As discussed in \cite{Farzan:2016wym}, the LFV process $\tau \to e \gamma$ is GIM suppressed and does not yield a strong bound.

\section{Flux of neutrinos from pion decay into $Z'$  \label{spectrum}}
In this section, we compute the flux of neutrinos from pion decay, $\pi^+ \to Z' \nu l_\alpha^+$ as well as from subsequent $Z'$decay $Z'\to \nu \bar{\nu}$. Similar formulas hold valid for $K^+\to Z' \nu l_\alpha^+$ with replacing $f_\pi \to f_K$, $m_\pi \to m_K$, and $\cos \theta_C \to \sin \theta_C$. Details of the calculation are outlined in the Appendix. In general, we can write the flux of neutrinos in the lab frame, $\phi(E_\nu)$ as
\begin{equation}\label{Eq.nuflux}
\phi(E_\nu)= \frac{1}{4 \pi L^2} \int_{E_\pi^{min}}^{E_\pi^{max}}dE_{\pi} P_\pi(E_\pi) (\frac{dN_\nu}{dE_\nu})_{lab} \frac{d \Omega_{r.\pi}}{d \Omega_{lab}},
\end{equation}
where  $L$ is the distance from the source to the  detector and $P_\pi(E_\pi)$ is the rate of the pion  injection in the lab frame.
  $(\frac{dN_\nu}{dE_\nu})_{lab}$ is the spectrum of the neutrino in the lab frame from the decay of a pion with an energy of $E_\pi$ and is  related to the spectrum of neutrinos in the rest frame of pion, $dN_\nu/dE_\nu|_{r.\pi}$, as
\be
(\frac{dN_\nu}{dE_\nu})_{lab}=(\frac{dN_\nu}{dE_{\nu}})|_{r.\pi} \frac{ \partial E_\nu|_{r.\pi}}{ \partial E_\nu|_{lab}}.
\ee
Setting the angle between the direction of neutrinos reaching the detector and the momentum of the pion beam to zero\footnote{Remember that the DUNE setup is going to be on-axis and the angle subtending the ND is $\sim O(3.5 ~{\rm m}/570 ~{\rm m})$.}, we can write $E_\nu|_{lab}=E_\nu|_{r.\pi}(1+v_\pi)\gamma_\pi$, in which $v_\pi$ is the pion velocity in the lab frame and $\gamma_\pi=(1-v_\pi^2)^{-1/2}$. Thus,
$$
\frac{\partial E_\nu|_{r.\pi}}{ \partial E_\nu |_{lab}}=\gamma_\pi (1-v_\pi).$$
Finally, $d \Omega_{r.\pi}/d\Omega_{lab}=(1+v_\pi)/(4(1-v_\pi)) \simeq \gamma_\pi^2$ takes care of focusing of the beam in the direction of the detector.

In the following, we compute the spectrum of neutrinos from $\pi \to e \nu_\alpha Z'$ and $Z' \to \nu \bar{\nu}$, neglecting the mass of the electron. We return to the case of $\pi \to \mu \nu Z'$ at the end of this section.
It is straightforward to show that the spectrum of the prompt $\nu$ in the rest frame of the pion is given by
\begin{equation}
\frac{d\Gamma(\pi\longrightarrow e\nu_\alpha Z')}{dE_\nu}|_{r.\pi}=\frac{f_{\pi }^2 g_{e\alpha}^2 G_F^2 \cos^2 \left(\theta _C\right)}{64 \pi ^3 m_\pi^2  m_{Z'}^2 (m_\pi-2 E_\nu)^2}E_\nu^2 \left(2 E_\nu m_\pi-m_\pi^2+m_{Z'}^2\right)^2 \left(-2 E_\nu m_\pi+m_\pi^2+2 m_{Z'}^2\right).
\end{equation}
Computing the spectrum of neutrinos for $Z'$ decay in the rest frame of the pion is slightly more complicated. The energies of $\nu$ and $\bar{\nu}$ in the rest frame of
$Z'$ are equal to $m_{Z'}/2$, but their energies in the rest frame of the pion depend on the angle between the direction of their emission and the direction of the $Z'$ momentum.
Moreover, the angular distribution of the neutrino and antineutrino (even in the rest frame of $Z'$) depends on the polarization of $Z'$. We therefore have to make the computation for each $Z'$ polarization separately.

In the rest frame of the pion, the differential decay rate of the pion to the electron, neutrino and $Z'$ with  polarization  perpendicular to the $Z'$ momentum $(\epsilon_1,\epsilon_2)$ and parallel $(\epsilon_3)$
to the $Z'$ momentum are, respectively,\be
\frac{d\Gamma(\pi\longrightarrow e\nu_\alpha Z')}{dE_{Z'}} \mid_{1,2}=\frac{ f_{\pi }^2 g_{e\alpha}^2 G_F^2 \cos^2 \left(\theta _C\right)}{96 \pi ^3 m_\pi}p_{Z'} \left(-2 E_{Z'} m_\pi+m_\pi^2+m_{Z'}^2\right)
\ee
and
\be
\frac{d\Gamma(\pi\longrightarrow e\nu_\alpha Z')}{dE_{Z'}} \mid_{3}=\frac{f_{\pi }^2 g_{e\alpha}^2 G_F^2 \cos^2 \left(\theta _C\right)}{96 \pi ^3 m_\pi  m_{Z'}^2}p_{Z'} \left(E_{Z'} m_\pi-m_{Z'}^2\right)^2.
\ee
Notice that, as expected, the decay into the longitudinal mode is proportional to $g_{e\alpha}^2/m_{Z'}^2$ and will be enhanced for  $m_{Z'}\ll m_\pi$.

The total decay rate of the $Z'\longrightarrow\nu_\alpha\bar{\nu}_\beta$ for all the polarizations is
equal to
$$
\Gamma(Z'\longrightarrow\nu_\alpha\bar{\nu}_\beta)=\frac{g^2_{\alpha\beta} m_{Z'}}{24\pi}.
$$
The conditions for $Z'$ decay before reaching the near and far detectors of DUNE
 ($\Gamma (m_{Z'}/E_{Z'}) L >1$) are, respectively,
 \be\label{Eq.8}
 g_{\alpha \beta}> 2\times 10^{-7} (10~{\rm MeV}/m_{Z'}) , \;   g_{\alpha \beta}> 4\times 10^{-9} (10~{\rm MeV}/m_{Z'}).
 \ee
The neutrino spectrum produced from  $Z^\prime$ decay with polarization $i$ in the rest frame of $Z'$ is given by
\be
(\frac{dN_\nu}{d\Omega})_{r.Z'}|_i= \frac{1}{\Gamma(Z'\longrightarrow \nu \bar{\nu})} \frac{d\Gamma(Z'\longrightarrow \nu \bar{\nu})}{d \Omega}|_i,
\ee
For $\epsilon_1$ and $\epsilon_2$, the normalized spectrum is
\be\label{1,2}
(\frac{dN_\nu}{d\cos\theta})_{r.Z'}\mid_{1,2}=\frac{3(1+\cos^2\theta) }{8},
\ee
and in the case of $\epsilon_3$ is
\be\label{3}
(\frac{dN_\nu}{d\cos\theta})_{r.Z'}\mid_{3}=\frac{3\sin^2\theta}{4},
\ee
where $\theta$ is the angle between the direction of the neutrino in the rest frame $Z'$ and the direction of $Z'$ emission in the rest frame of the pion.
As expected $\sum_{i=1}^3 (dN_\nu/d\cos \theta)_{r.Z'}$ is independent of $\theta$.
From Eqs. (\ref{1,2}) and (\ref{3}), we observe that   the angular distribution is invariant  under $\theta \to \pi-\theta$ which means that the    neutrinos and antineutrinos from the $Z'$ decay have the same angular distribution. This is rather counter-intuitive,
because angular momentum conservation combined with the fact that (anti)neutrinos are (right-)left-handed implies that if the spin of $Z'$ is in, {\it e.g.,} the state of $|l=1,m=+1 \rangle$, antineutrinos (neutrinos) will be emitted parallel  (antiparallel) to the spin direction. Notice, however, that none of the 1, 2 and 3 polarization corresponds to   $|l=1,m=+1 \rangle$ or  $|l=1,m=-1 \rangle$. In fact, the 3 polarization corresponds to   $|l=1,m=0 \rangle$ and the 1 and 2 polarizations are linear combinations of   $|l=1,m=\pm 1 \rangle$. Thus, angular momentum conservation implies that the decays of each  of these polarizations lead to the emission of both parallel and antiparallel neutrinos and antineutrinos as reflected in Eqs (\ref{1,2}) and (\ref{3}).
 Remembering that the energy of $\nu$ (or $\bar{\nu}$) in the rest frame of $Z'$ is $m_{Z'}/2$, its energy in the pion rest frame will be
\be \label{cos} E_\nu|_{r.\pi}= \frac{m_{Z'}}{2}\gamma_{Z'}(1+v_{Z'} \cos \theta)\ee
in which  $v_{Z'}$ is the velocity of the $Z'$ in the rest frame of the pion, $v_{Z'}=(1-m_{Z'}^2/E_{Z'}^2)^{1/2}$, and $\gamma_{Z'}=(1-v_{Z'}^2)^{-1/2}$.
In the rest frame of the pion, the spectrum of the neutrino produced from $Z'$ decay is then
\be
(\frac{dN_\nu}{dE_\nu})_{r.\pi}\mid_{i}=(\frac{dN_\nu}{d\cos\theta})_{r.Z'}\mid_{i} \frac{d\cos\theta|_{r.Z'}}{dE_\nu|_{r.\pi}}= (\frac{dN_\nu}{d\cos\theta})_{r.Z'}\mid_{i} \frac{2}{E_{Z'} v_{Z'}}.
\ee
We already noticed that $(d N_\nu/d\cos \theta)_{r.Z'}=(d N_{\bar{\nu}}/d\cos \theta)_{r.Z'}$, so in the rest frame of the pion, the energy spectrum of neutrinos and antineutrinos from $Z'$ decay will be the same.
The total neutrino spectrum from $Z'$ decay in the rest frame of pion is given by
\be
(\frac{dN_\nu}{dE_\nu })_{r.\pi}^{Z'~ decay}= \sum_i\int_{E_{Z'}^{min}}^{E_{Z'}^{max}} dE_{Z'}    \frac{dN_{Z'}}{dE_{Z'}}\mid_{i} (\frac{dN_\nu}{dE_\nu})_{r.\pi}\mid_{i}
\ee
in which $i$ refers to the $Z'$ polarization, $E_{Z'}^{min}=E_\nu+m_{Z'}^2/(4E_\nu)$, $E_{Z'}^{max}=(m_\pi^2+m_{Z'}^2)/(2m_\pi)$, and
\be
\frac{dN_{Z'}}{dE_{Z'}}\mid_{i}=\frac{1}{\Gamma^\pi_{total}}\frac{d\Gamma(\pi\longrightarrow e\nu Z')}{dE_{Z'}}\mid_{i}.
\ee

A similar consideration holds for the decay $\pi^+ \to \mu^+ Z'\nu$ with the difference that the mass of $\mu^+$ cannot be neglected. To avoid cluttering, we will not show  the formulas here. Notice, however, that, because of phase space, the decay into muon is significantly suppressed. In  fact for $m_{Z'}>5$ MeV, $\Gamma(\pi \to \mu \nu_\alpha Z')/ \Gamma(\pi \to e \nu_\alpha Z')$ is smaller than 0.055$|g_{\mu \alpha}|^2/|g_{e \alpha}|^2$ and reaches $<0.01|g_{\mu \alpha}|^2/|g_{e \alpha}|^2$ from $m_{Z'}> 20$ MeV.

\section{Light $Z'$ and the DUNE near detector  \label{DUNEres}}
\begin{figure}
\begin{center}
        \includegraphics[width=0.6\textwidth]{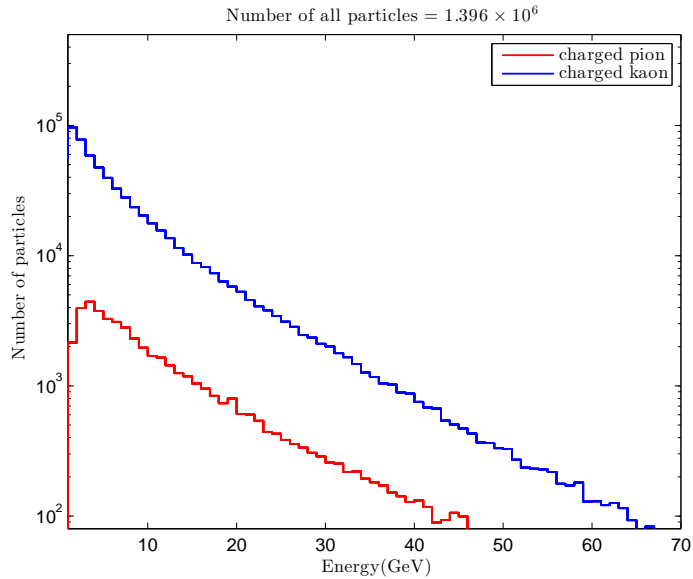}
 \end{center}
\caption{\label{pionkaonspectrum} The spectrum of the charged pion and kaon in the beginning of the decay pipe in the DUNE experiment, taking the total  number of all particles equal to $1.396 \times 10^6$
\cite{Leo}.}
\end{figure}

\begin{figure}
\begin{center}
        \includegraphics[width=0.6\textwidth]{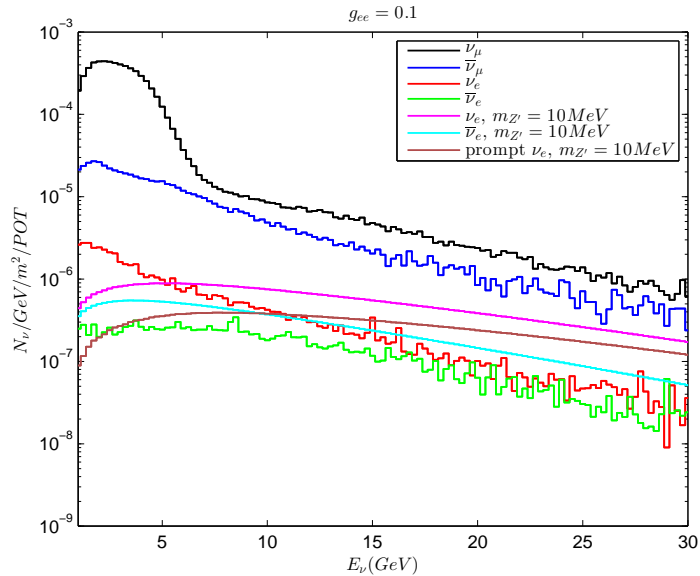}
 \end{center}
\caption{\label{nu_flux} Fluxes of neutrinos at the near detector of DUNE  for different flavors within the standard model \cite{DUNEflux} in the $\pi^+$ mode. 
The cyan, brown, and magenta lines, respectively, correspond to  $\bar{\nu}_e$ from $Z'$ decay, prompt $\nu_e$ from $\pi^+ \to e^+ \nu_e Z'$, and total $\nu_e$ flux from
$\pi^+ \to e^+ \nu_e Z'$ plus that from subsequent $Z' \to \nu_e\bar{\nu}_e$ for  $g_{ee}=0.1$ and $m_{Z'}$=10 MeV.}
\end{figure}

{
In this section, we discuss how the data from the DUNE near detector can be used to extract information on the coupling of neutrinos to $Z'$.
To compute the flux of neutrinos from the new pion and kaon decay mode, we have used the formalism developed in Sec. \ref{spectrum}. The energy spectra of the charged pion and kaon are  demonstrated in Fig. \ref{pionkaonspectrum} \cite{Leo}, which correspond to  the 120 GeV mode and the total number of all particles equal to  $1.396\times 10^6$. Figure \ref{nu_flux} shows the predicted flux of neutrinos of different flavors from the $\pi^+$ decay  within the SM \cite{DUNEflux}  and compares it with the  $\nu_e$ and $\bar{\nu}_e$ spectra from new physics with $Z'$  of 10 MeV  mass and a coupling of $g_{ee}=0.1$. Such a large value of $g_{ee}$ is already ruled out, but we have chosen this value for illustrative purposes.  For smaller values of $g_{ee}$, the flux will be just scaled by $g_{ee}^2$.} The
cyan curve shows the electron {(anti)neutrino} spectrum coming from $Z^{\prime}$ decay.  {The brown curve shows the electron neutrino coming from $\pi^+$ decay.}
 The magenta curve shows the electron neutrino spectrum coming from both $Z^\prime$ and $\pi^+$ decays. Remembering that $\nu$ and $\bar{\nu}
 $ from the $Z'$ decay have the same  energy spectrum,  the difference between the purple and cyan curves gives the spectrum of neutrinos produced from  the pion decay, $\pi^+ \to e^+ Z'\nu$,  {\it i.e.,} {the brown curve}.

 To compute the number of events, we  have taken the neutrino cross section from Refs. \cite{Messier:1999kj,Paschos:2001np}.
 For the analysis, we consider {5+5} years of data taking  with 1.5$\times 10^{21}$ POT/year in {both} $\pi^+$ {and $\pi^-$}  decay modes, for neutrinos with the energies from 1 to 30 GeV, dividing to 29 equal size bins.
Notice that, although for $g_{ee}\ll 0.1$, the number of signal events will be much smaller than that of the background (see Fig. \ref{nu_flux}), with this amount of POT, as long as $g_{ee}\sim 10^{-3}$, the square root of background events at each bin ({\it i.e.,} the statistical uncertainty of the background) can be smaller than the number of signal events at that bin. As a result, the results can be sensitive to such small values of the coupling.
For simplicity in our analysis, we consider perfect energy resolution and efficiencies or acceptance  for the $\nu_e$ and $\bar{\nu}_e$ at the DUNE near detector. In fact, we repeated the analysis, setting the efficiencies of the $\nu_e$ and $\nu_\mu$ detection to 80$\%$ and 85$\%$ \cite{Adams:2013qkq}, respectively, and simultaneously increasing the total flux by a factor of 1.25.  We then found that the results are unchanged. The near detector of DUNE will measure the absolute flux to 2.5$\%$ precision, via neutrino-electron scattering \cite{Acciarri:2015uup}.
For statistical inference, we have used the chi-squared method defined as
\begin{equation}\label{chi2}
\chi^2=
\sum_{bins}\frac{[N_{th}^i(1+\eta)-N_{exp}^i]^2 }{\sigma_i^2}+\frac{\eta^2}{(\Delta \eta)^2},
\end{equation}
in which $N_{th}^i$ is the theoretical prediction for the number of events in each bin, assuming a certain value of $g_{e\alpha}$ and $m_{Z'}$. $N_{exp}^i$ is the forecast for median number of events in each, assuming a certain true value for $m_{Z'}$ and coupling.
 $\eta$ is the nuisance parameter that takes care of the normalization uncertainty of $\Delta \eta =2.5\%$.
The details of the near detector are under discussion, however, we assume an 84 ton fiducial mass of the liquid argon time projection chamber (LArTPC) detector and multi-purpose tracker at 575 m from the source.

\begin{figure}
\begin{center}
\includegraphics[width=0.6\textwidth]{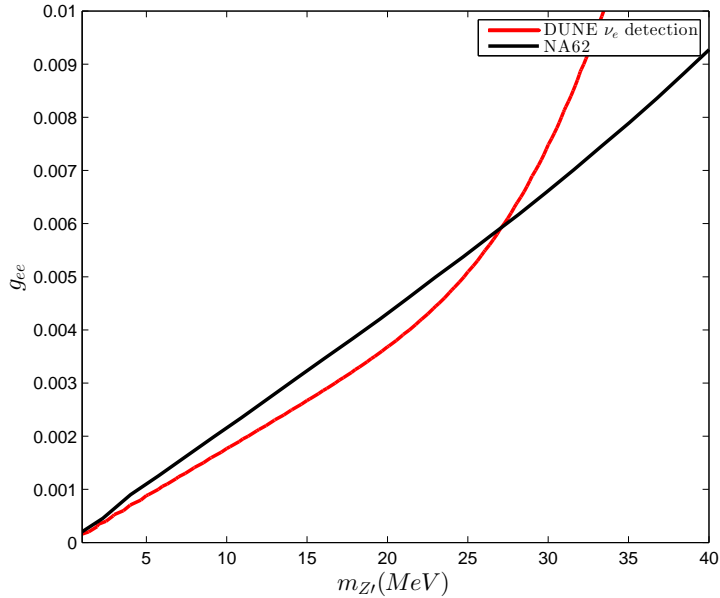}
 \end{center}
\caption{\label{constrain} The upper bound on $g_{ee}$ vs. $m_{Z^{\prime}}$ at 90\% C.L. The red curve shows the DUNE constraint on $g_{ee}$ by $\nu_e$ detection
  after five years of data taking with 1.5$\times 10^{21}$ POT/year  in both $\pi^+$ and $\pi^-$ modes.  The black line shows the NA62 current constraint on $g_{ee}$ \cite{Bakhti:2017jhm}.}
\label{result}
\end{figure}

 Figure \ref{result} shows the 90$\%$ C.L.  constraints on the $g_{e e}$ versus $m_{Z'}$ setting all the true values $g_{\alpha \beta} = 0$. We have considered neutrinos from both kaon and pion decay. The best current constraint on $\sqrt{\sum_\alpha g_{e\alpha}^2}$ which comes from the kaon decay measurement at NA62 experiments  \cite{Bakhti:2017jhm}, is also shown in Fig. \ref{result}. Comparing with current constraints, the near detector of DUNE will constrain the coupling stronger for $m_{Z'}<30$ MeV. Above 30 MeV, the current bound from NA62 is stronger than the reach of  DUNE because, while
for heavy $Z'$, $\pi\to Z'\nu e$ is suppressed due to the phase space, $m_{Z'}$ is still much lighter than $K$ and therefore $K\to Z'\nu e$  is not suppressed. For $m_{Z'}<25$ MeV, the neutrinos from $\pi^+$ at DUNE dominate over those from $K^+$. That is for $m_{Z'}<25$ MeV, even if we neglected the $K^+$ flux, we would obtain the same bound from DUNE. The contribution from $K^+$ becomes important only for $m_{Z'}>30$ MeV where the NA62 bound is stronger. As a result, across the $m_{Z'}$ range that we expect DUNE to improve the NA62 bounds, the results will not suffer from uncertainties in computing the $K^+$ flux.

\begin{figure}
\begin{center}
\includegraphics[width=0.6\textwidth]{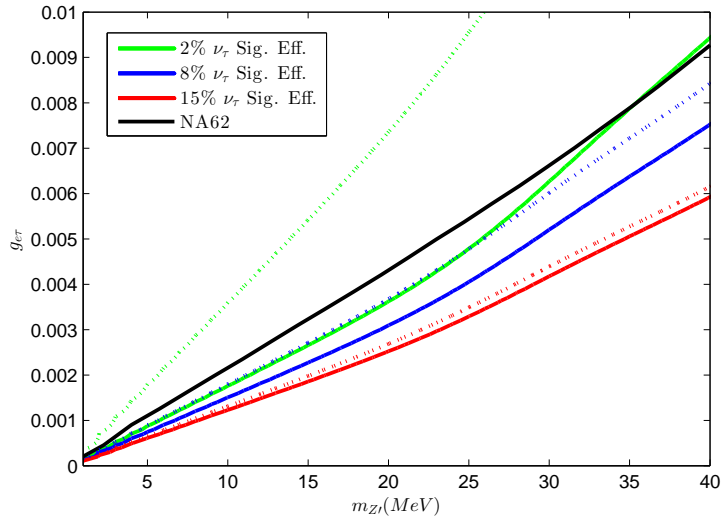}
 \end{center}
\caption{\label{NOE} The upper bound on $g_{e \tau}$ vs. $m_{Z^{\prime}}$ at 90 \% C.L., setting  the rest of $g_{\alpha\beta}=0$. The  dashed  colored lines show the constraint on $g_{e \tau}$ using only $\nu_\tau$ and $\bar{\nu}_\tau$ signals with detection efficiencies of 2\%, 8\% and 15\% for 1.5$\times 10^{21}$ POT/year after five years of data taking in each mode.  The corresponding solid lines show the constraints from combining all $\nu_e$, $\bar{\nu}_e$, $\nu_\tau$ and $\bar{\nu}_\tau$ signals.   The black curve shows the current bound from NA62 \cite{Bakhti:2017jhm}.}
\end{figure}

\begin{figure}
\begin{center}
\subfigure{\includegraphics[width=0.45\textwidth]{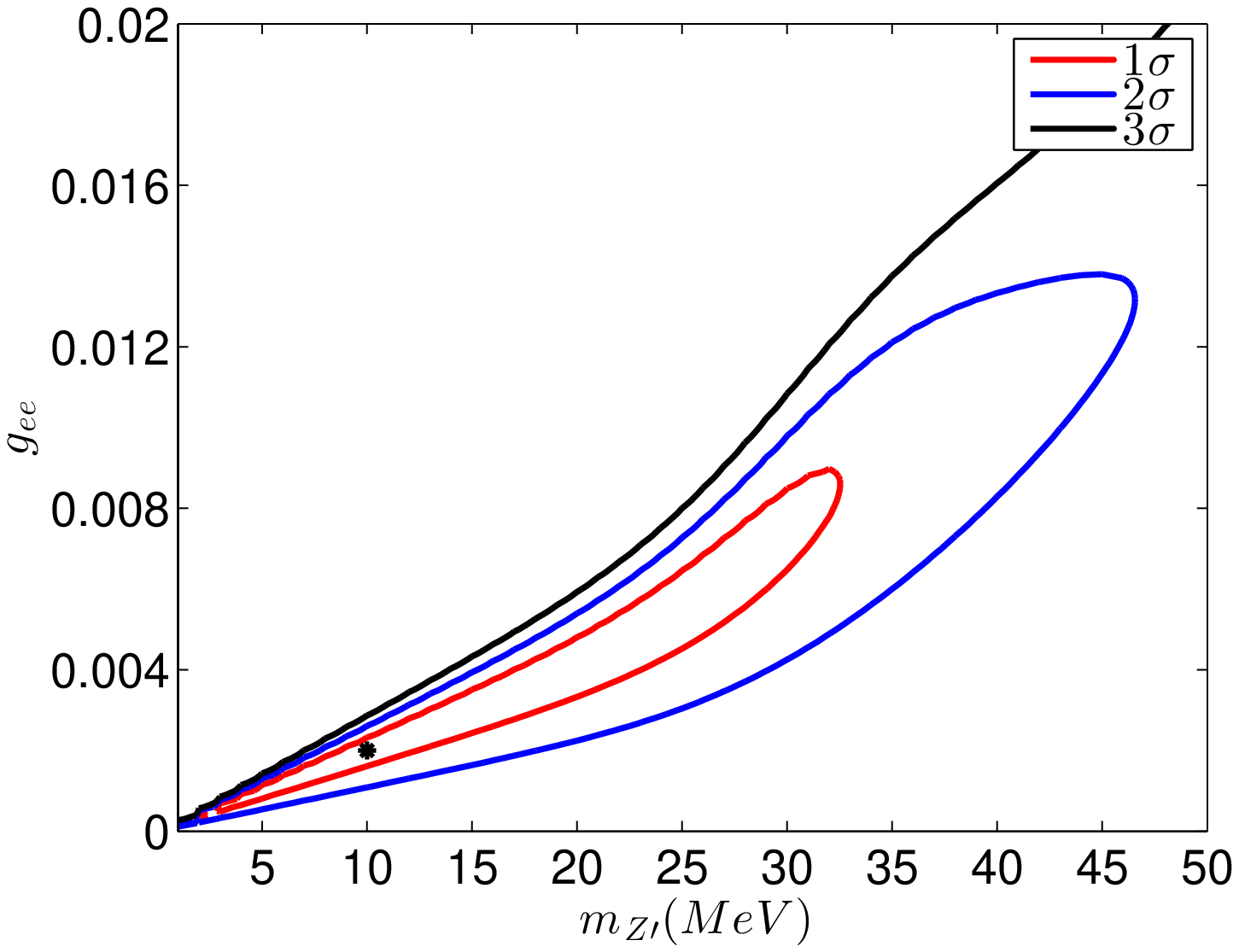}}
\subfigure{\includegraphics[width=0.45\textwidth]{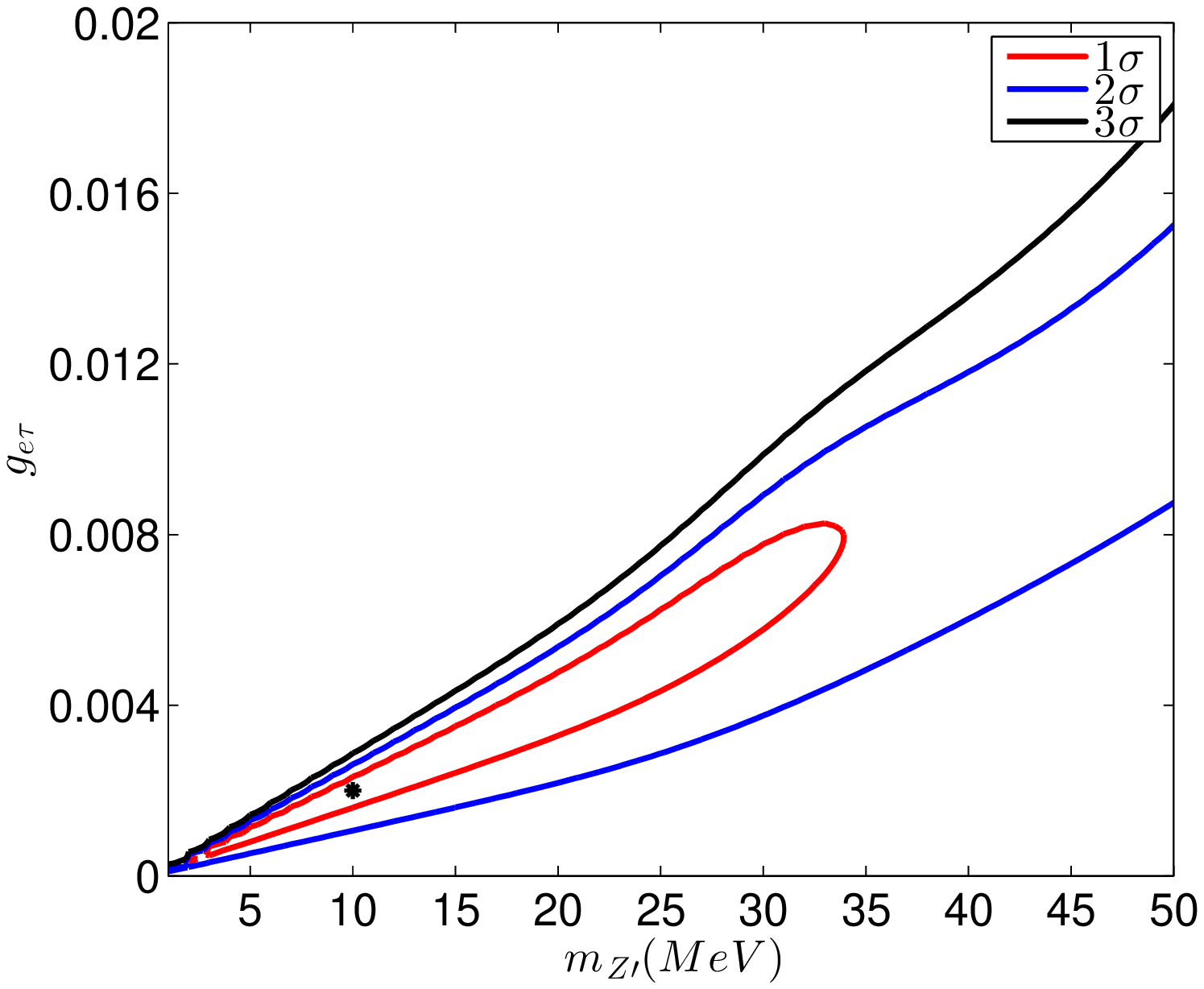}}
 \end{center}
\caption{\label{constrain1}   Discovery potential of the near detector of DUNE. The black dot shows the assumed true values for $m_{Z'}$ and $g_{ee}$ in the left panel ($g_{e\tau}$ in the right panel). The contour plots are 1$\sigma$, 2$\sigma$, and 3$\sigma$  solutions from the DUNE near detector after ten years of data taking with  1.5 $\times 10^{21}$ POT/year in  the $\pi^+$ mode. To extract the value  of $g_{e\tau}$ in the right panel, we have assumed 8\% efficiency and have combined  the information from all $\nu_e$, $\bar\nu_e$, $\nu_\tau$, and $\bar\nu_\tau$ signals.}
\end{figure}

\begin{figure}

\centering{
\subfigure[]{\label{a} \includegraphics[width=0.40\textwidth]{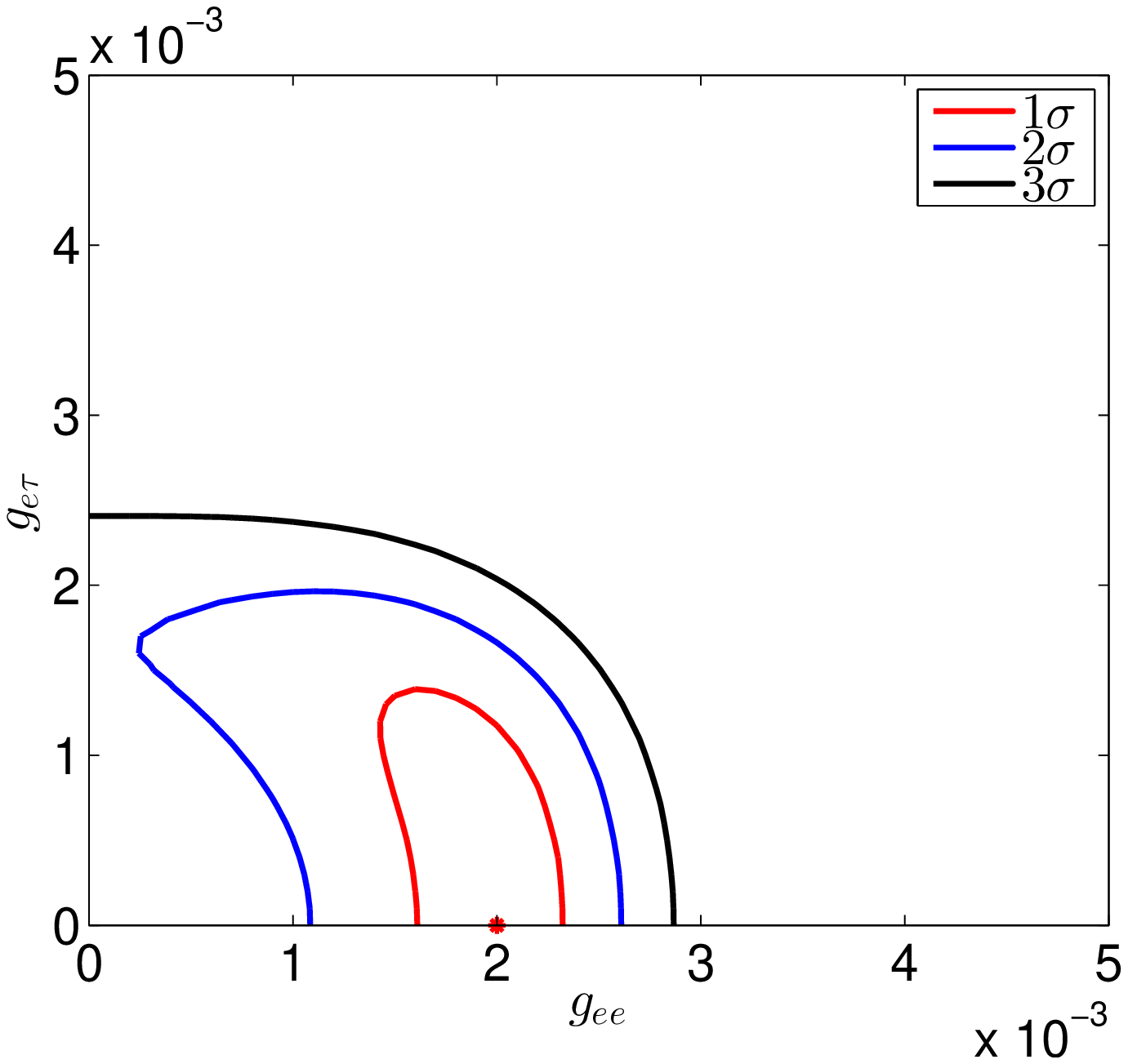} \qquad }
\subfigure[]{\label{b}\includegraphics[width=0.40\textwidth]{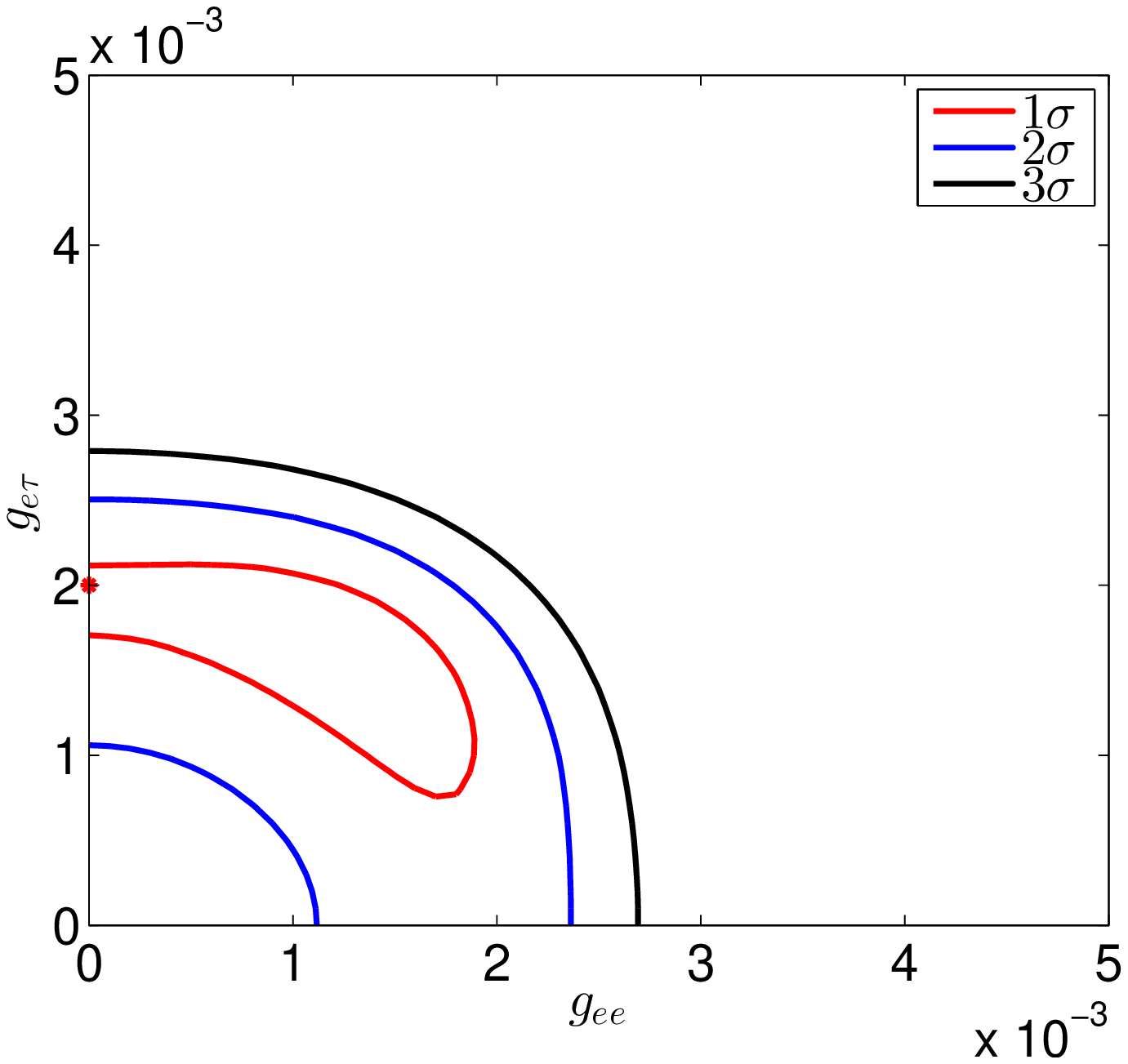}  \qquad}
\subfigure[]{\label{c} \includegraphics[width=0.40\textwidth]{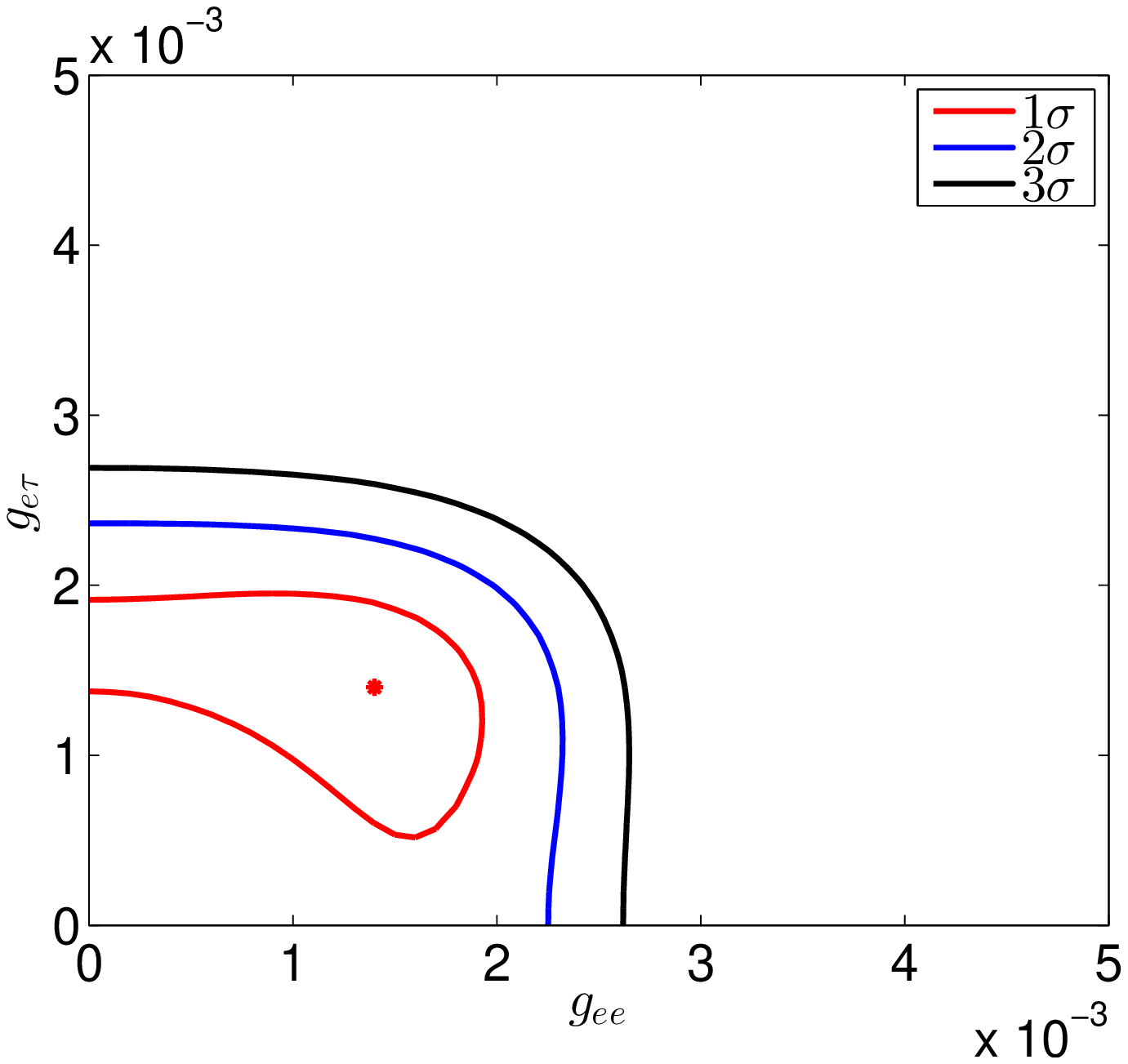} \qquad}
\subfigure[]{\label{d}\includegraphics[width=0.40\textwidth]{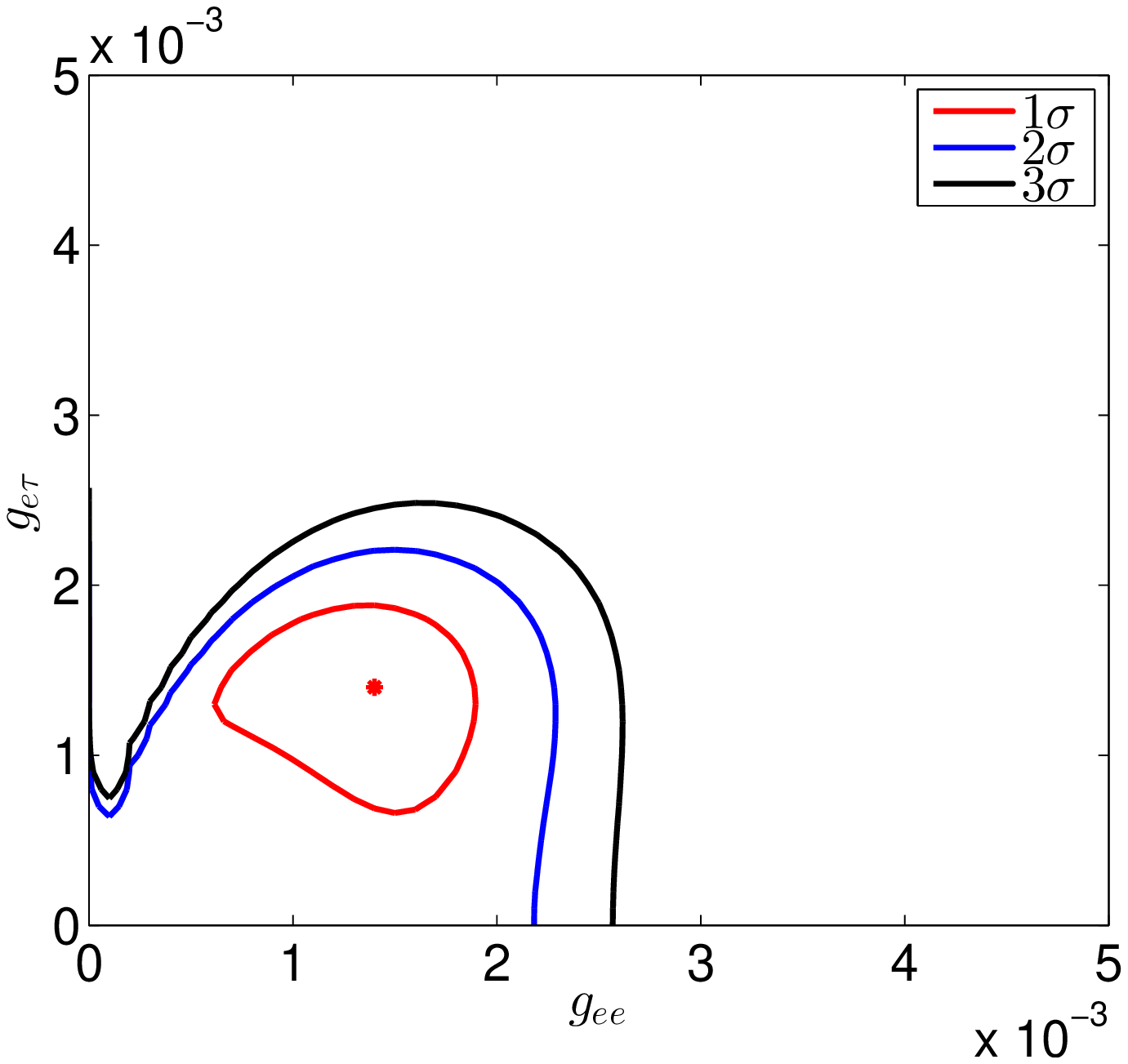} \qquad }
}
\caption{\label{constrain1} The capability of the DUNE ND in reconstructing the flavor structure of $g_{\alpha \beta}$.  The true value of $m_{Z'}$ is taken to be 10 MeV. The red dot in each plot shows the assumed true value of $g_{ee}$ and $g_{e\tau}$. In panels (a,b,c), the rest of the coupling components are set to zero, but in (b), we take $g_{\tau \tau}=g_{e\tau}^2/g_{ee}$. The curves show the 1$\sigma$, 2$\sigma$ and 3$\sigma$ solutions fixing $m_{Z'}$ to 10 MeV after five years of data taking for 1.5 $\times 10^{21}$ POT/year in each mode with 8 \% $\nu_\tau$ detection efficiency. }
\end{figure}

 As is explained in the conceptual design of DUNE \cite{Acciarri:2015uup}, both the near and far detectors will be sensitive to the detection of $\nu_\tau$. In the case of nonzero $g_{e \tau}$, $\bar{\nu}_\tau$ and $\nu_\tau$ are produced via $\pi^+ \to e^+ Z' \nu_\tau$ followed by $Z' \to \nu_e \bar{\nu}_\tau$ and  $Z' \to  \bar{\nu}_e {\nu}_\tau$.  Thus, the near detector of DUNE has the potential to constrain $g_{e \tau}$ by $\nu_\tau$ and $\bar{\nu}_\tau$ detection.  The charged current interactions of $\nu_\tau$ ($\bar{\nu}_\tau$) at the detector can lead to $\tau$ ($\bar{\tau}$) production. Three decay modes of $\tau$ (and $\bar{\tau}$) can be used for reconstructing $\nu_\tau$ (and $\bar{\nu}_\tau$):
(i) $\tau^- \rightarrow e^- \bar{\nu} _e \nu _\tau$ with a branching ratio of $17.8 \%$,  (ii) $\tau ^- \rightarrow \pi ^- \nu _\tau (n \pi ^0)$ with a branching ratio of $49.8 \%$ and (iii) $\tau ^- \rightarrow  \pi^- \pi^+ \pi^- \nu _\tau (n \pi^0) $ with a branching ratio of $15.2\%$. We have considered only hadronic decay channels in our analysis. To our best knowledge, the background for $\nu_\tau$ and the detection efficiency at DUNE are not still known. To carry out our analysis, we rely on the estimates that have been made for previous setups, keeping in mind that superior particle identification power at DUNE will lead to better estimating the $\nu_\tau$ detection efficiency and more suppressed background.
 In Ref. \cite{Conrad:2010mh}, the backgrounds for atmospheric $\nu_\tau$ are estimated for a LARTPC detector. Notice that the atmospheric neutrino flux is isotropic. In the beamed experiments such as NOMAD or DUNE, kinematic cuts can help to significantly  reduce the background.
The main source of the background for $\nu_\tau$ detection via a hadronic channel is the neutrino neutral-current interaction.  Since the efficiency of GeV scale $e$ and $\mu$ identification at the LAR neutrino detectors is very high \cite{Conrad:2010mh}, the background from the charged-current interactions of $\nu_\mu / \bar{\nu}_\mu $ and  $\nu_e / \bar{\nu}_e$ is negligible for $\nu _\tau$ detection.
In Ref. \cite{Astier:2000ng}, the backgrounds for $\nu_\tau$ have been obtained for NOMAD experiment using kinematic cuts on the momenta of $\bar{\nu}_\tau$ and $\nu _\tau$. For the strongest kinematic cuts which were designed to reduce background from both neutral current and charged current interactions, the signal efficiency  for hadronic mode at NOMAD was 2\% \cite{Astier:2000ng}  and the background rejection for neutral current in hadronic channel was $2\times10^{-5}$. At DUNE, the kinematic cuts can be less stringent because  only neutral currents induce background. In addition, DUNE near detector is more sensitive to neutron and pion detection than the NOMAD experiment, further enabling reduction of the background. In the case of neutron detection, as is mentioned in the conceptual design of DUNE \cite{Acciarri:2015uup}, the future experiment CAPTAIN will be able to determine the response of LARTPC to neutrons with higher accuracy. Thus, it is reasonable to take an efficiency for $\nu_\tau$ detection higher than 2 \%. In our analysis, we take an efficiency of $2 \times 10^{-5}$ in suppressing the background from neutral current events (the same as NOMAD), and we take various values of the $\nu_\tau$ detection efficiency between 2 \% and 15 \%. One should, however, bear in mind that the NC background reduction at DUNE can also be much better than that at NOMAD.

Figure 4 shows the bound that the DUNE ND can set on $g_{e\tau}$ if there is no new physics ({\it i.e.,} if $g_{\alpha \beta}=0$) and compares it to the bound from NA62 on $(\sum_\alpha |g_{e\alpha}|^2)^{1/2}$. While to draw the dashed lines only $\nu_\tau$ and $\bar{\nu}_\tau$ signals are invoked, to draw the solid lines all the $\nu_e$, $\bar{\nu}_e$, $\nu_\tau$ and $\bar{\nu}_\tau$ signals are used. The green, blue, and red lines, respectively, correspond to 2 \%, 8 \%, and 15\% efficiencies in $\nu_\tau$ and $\bar{\nu}_\tau$ detection. As seen, the bound significantly improves with increasing the detection efficiency. Even with a 2 \% efficiency, the combined electron- and tau-neutrino signals lead to a bound on $g_{e\tau}$ stronger than that from NA62.

Figure 5 shows the discovery potential of the DUNE ND assuming that values of couplings are slightly below the current bound from NA62. At each panel only one $g_{\alpha \beta}$ is taken to be nonzero. As seen from the figure, DUNE can determine the ratio $g_{ee}/m_{Z'}$ or  $g_{e\tau}/m_{Z'}$ at 2 $\sigma$ C.L.  For the values of coupling considered, $g_{e\tau}\to 0$ (at a finite $m_{Z'}$) is ruled out at 2 $\sigma$ C.L. As seen also at 2 $\sigma$ C.L., an upper bound on $m_{Z'}$ can be derived but $m_{Z'},g\to 0$ (with fixed $g/m_{Z'}$) cannot be ruled out. This is because as $m_{Z'}\to 0$, the rate of $\pi^+ \to e^+ \nu Z'$ increases as $g^2/m_{Z'}^2$ without any particular feature in the energy spectrum.

Figure 6 shows how the DUNE ND can reconstruct the flavor structure of the coupling. To draw Fig. 6-a, 6-b and 6-c, only the $ee$ and $e\tau$ components are allowed to be nonzero, but to draw Fig. 6-d, we take $g_{\tau \tau}=g_{e\tau}^2/g_{ee}$ as predicted by the model presented in Sec. II. The $\nu_\tau$ detection efficiency is taken to be $8 \%$. As seen from Figs 6-a and 6-b, for the assumed values of couplings, the DUNE ND can tell us whether  the signal comes from nonzero $g_{ee}$  or from nonzero $g_{e \tau}$  at 2 $\sigma$ C.L. This is the great advantage of the ND DUNE over the kaon decay experiments such as NA62, which can determine only $g_{ee}^2+g_{e\mu}^2+g_{e\tau}^2$ and originates from the fact that the DUNE ND can detect
the final neutrino states.
Notice that for drawing these figures, we have assumed that we know the true value of $m_{Z'}$. That is in computing $\chi^2$, we have set $m_{Z'}$ in both $N_{exp}^i$ and $N_{th}^i$ equal to 10 MeV. Of course in reality, $m_{Z'}$ will be unknown. However, the lack of knowledge of $m_{Z'}$ does not ruin the flavor reconstruction power of DUNE ND because it comes from the ratio of the $\nu_\tau +\bar{\nu}_\tau$ signal to the $\nu_e +\bar{\nu}_e$ signal rather than the absolute number of events. We examined this by taking the ``true" value of $m_{Z'}=10$ MeV and the ``assumed
" value of $m_{Z'}$ in $N_{th}^i$ equal to 25 MeV. As expected the contours turned up to be similar to those shown in  Fig. 6 but scaled up by the ratio of the two true and assumed values of $m_{Z'}$.

In the discussion above, we focused on $Z'$ with a mass above a few MeV. For keV$<m_{Z'}<$ a few MeV, the coupling of $Z'$ to neutrinos is severely constrained by cosmological observations \cite{Huang:2017egl}, as $Z'$ can appear as extra relativistic degrees of freedom. For smaller $Z'$, the bound from cosmology relaxes, so the bounds from the meson decay can become relevant again. Notice that the strong bounds from stellar cooling for  $m_{Z'}\sim$ keV \cite{Redondo:2013lna,Hardy:2016kme} and from the fifth force searches for $m_{Z'}<10$ eV \cite{Bordag:2001qi} do not apply to our scenario, because our $Z'$ does not couple to the matter fields. For $m_{Z'}<$keV, the bound from NA62 can be straightforwardly  extrapolated as

$$    \sqrt{\sum_\alpha |g_{e\alpha}|^2}<2.23\times 10^{-10} \frac{m_{Z'}}{{\rm eV}}.$$

Extrapolating the bound from DUNE is less straightforward, because, for smaller $Z'$ and $g_{\alpha \beta}$, the $Z'$ particles decay only after passing the near detector [see Eq. (\ref{Eq.8})]. The bound will therefore come only from searching for  the prompt $\nu_\alpha$ from the pion decay. Taking this fact into account, we found that, for $m_{Z'} <0.1$ MeV, the DUNE near detector can set the following bounds:

$$ g_{ee}< 3.0\times 10^{-10}   \frac{m_{Z'}}{{\rm eV}} \  \ {\rm and} \ \    g_{e\tau}<1.9\times 10^{-10} \frac{m_{Z'}}{{\rm eV}} , $$
where for $\nu_\tau$ detection, an efficiency of 8 \% was assumed.

Notice that so far  only $ee$ and $e\tau$ components are discussed. If only $g_{\tau \tau}$ is nonzero, we do not expect any signal at the DUNE ND as $\pi \to \tau Z' \nu_\tau$ is kinematically forbidden. However, the $g_{e \mu}$, $g_{\mu \mu}$, and $g_{\mu \tau}$ components can lead to a signal. If only $g_{\mu \tau}$ and/or $g_{\mu\mu}$ are nonzero, the new decay mode for the pion will be $\pi\to \mu Z'\nu$, which is suppressed due to the phase space suppression; {\it i.e.,} $(m_\pi-m_\mu)/(2m_\pi) \ll 1$. However, with $g_{e\mu}$, we can have $\pi\to e Z' \nu$ and subsequently $Z'\to \nu_e \bar{\nu}_\mu, \ \nu_\mu \bar{\nu}_e$.  Because of the large background for $\nu_\mu$ and $\bar{\nu}_\mu$, we should rely only on $\nu_e$ and $\bar{\nu}_e$ detection for reconstructing $g_{e \mu}$. If we allow both $g_{ee}$ and $g_{e\mu}$ to be nonzero, a degeneracy between them will emerge.
However, the combination of $g_{e\mu}$ and $g_{ee}$ that DUNE is sensitive to is different from the combination that the kaon or pion decay experiments such as NA62 or PIENU can probe. Thus, in the future, combining positive signals from the two types of experiments may help to break the degeneracy.  DUNE can itself break the degeneracy between $g_{e\mu}$ and $g_{e\tau}$ by looking for the $\nu_\tau$ signal.

\section{Summary and Discussion \label{Summary}}

We have studied the effects of the couplings of the form $g_{\alpha \beta} \bar{\nu}_\alpha \gamma^\mu \nu_\beta Z_\mu^\prime$ with $m_{Z'}\sim {\rm few}~{\rm M}e{\rm V}-{\rm few}~10~{\rm M}e{\rm V}$ at  the near detector of DUNE. At the source, $Z^\prime$ can be produced via three-body decay of the charged mesons, along with an electron (or a muon) and a neutrino. Subsequently, $Z^\prime$ decays into a neutrino antineutrino pair before reaching the detector. The produced neutrinos can be detected at the near detector of DUNE.

The detection of $\nu_\mu$ and $\bar{\nu}_\mu$ signal suffers from a large background so we focus on $g_{ee}$ and $g_{e \tau}$ elements that lead to $\nu_e$, $\bar\nu_e$,  $\bar\nu_\tau$, and $\nu_\tau$ signals. Focusing only on the $g_{ee}$ component, we find that with $\sim 10^{22}$ POT, the ND of DUNE can improve the bound on $g_{ee}$ from NA62 for $m_{Z'}<30$ MeV. For heavier $Z'$, the decay mode $\pi \to Z' e \nu$ will be suppressed by phase space, so the NA62 experiment whose source is the kaon decay sets a stronger bound. To extract information on $g_{e\tau}$, all the ${\nu}_e$,  $\bar{\nu}_e$,  $\bar{\nu}_\tau$, and ${\nu}_\tau$
signals can be invoked. As seen in Fig. 4, the bound that can be derived on $g_{e \tau}$ strongly depends on the efficiency of the $\nu_\tau$ detection. Even with a 2 \% efficiency, combining
  the ${\nu}_e$, $\bar{\nu}_e$, ${\nu}_\tau$,  and $\bar{\nu}_\tau$
signals leads to a bound on $g_{e \tau}$ better than what can be derived from NA62 for the entire $m_{Z'}$ range probed. By improving the efficiency up to 8 \% (which considering the superb particle identification capabilities of DUNE, sounds feasible),  the bound can be significantly improved.

 We also  investigated the discovery potential of DUNE for nonzero $g_{ee}$ and/or $g_{e\tau}$. We found that, for a given true value of $m_{Z'}$, the ND of DUNE can set an upper bound on $m_{Z'}$; however, in the $g_{ee}$ (or $g_{e\tau}$) and $m_{Z'}$ plane,  the solution contours are elongated along $g_{e \alpha}/m_{Z'}=cte$ lines as $g_{e\alpha}\to 0$. This behavior originates from two facts: (i) for
 $m_{Z'}\to 0$, the longitudinal component of $Z'$ leads to a $g^2/m_{Z'}^2$ behavior in the rate of $\pi \to e Z' \nu$; (ii) for $m_{Z'}\to 0$, no special feature is expected.

While NA62 can only constrain $\sum_{\alpha \in \{ e , \mu, \tau\}} |g_{e \alpha}|^2$, the ND of DUNE has the advantage of determining $g_{e\tau}$ and $g_{ee}$, separately by studying the $\nu_e$ and $\nu_\tau$ signals.  We have studied this possibility and the results are shown in Fig. 6.

 The interactions that we are discussing can lead to a non-standard neutrino neutrino interaction with immediate consequences for supernova neutrinos   \cite{Das:2017iuj,Dighe:2017sur}. The effective potential is given by the ratio $g^2/m_{Z'}^2$
 which is a combination that can be well determined by the DUNE ND. As demonstrated in Fig. 5, if $(g_{ee}/m_{Z'})^2$ or $(g_{e\tau}/m_{Z'})^2$ are of the order of $4 \times 10^{-2} {\rm GeV}^{-2}\sim 4000 G_F$, the value of $g/m_{Z'}$ can be probed by the DUNE ND. If a positive signal is found, it will have a distinct observable effect on supernova neutrinos, so such information from the DUNE ND is going to be an invaluable  and unique input for analyzing supernova neutrinos.

 \subsection*{Acknowledgments}
We are grateful to Leo Bellantoni, Paul LeBrun and Laura Fields for providing us with information on the flux of charged mesons at the source of DUNE.
We also thank  C. Eilkinson, A. Ereditato, S. Pascoli, B. Choudhary, W. Rodejohann and A. Sousa for useful discussions and encouragement.
This project has received funding from the European Union's Horizon 2020 research and innovation programme under the Marie Sk\l{}odowska-Curie grant agreement No.~674896 and No.~690575.
Y.F. is also grateful to the ICTP associate office and to IFIC, Valencia University for warm and generous hospitality.
P.B. thanks MPIK for their kind hospitality and support. This work is supported by Iran Science Elites Federation Grant No. 11131.

\subsection*{Appendix:  COMPUTING THE NEUTRINO SPECTRUM \label{appendix}}
In this Appendix, we demonstrate the  details of the calculation of the neutrino spectrum from three-body decay of the pion ($\pi\longrightarrow e\nu_\alpha Z'$) and subsequent $Z'$ decay ($Z'\longrightarrow \nu \bar{\nu}$).
As discussed in Ref. \cite{Bakhti:2017jhm}, the rate of the new decay mode of the pion is given by
\begin{equation}\label{decayrate}
\Gamma(\pi\longrightarrow l\nu Z')=\frac{1}{64\pi^3m_\pi}\int_{E_{Z'}^{min}}^{E_{Z'}^{max}}\int_{E_\nu^{min}}^{E_\nu^{max}}  dE_{Z'}  dE_\nu \sum_{spins}\vert {\cal M} \vert^2.
\end{equation}
Neglecting the neutrino mass, the integration limits are given by
$$E_{Z'}^{min}=m_{Z'},~~~~~~~~~~~~~~~~~~~~~~~~~~~~~~~~~~~~ E_{Z'}^{max}=\frac{m_\pi^2+m_{Z'}^2-m_l^2}{2m_\pi},$$
$$E_\nu^{min}=\frac{m_\pi^2+m_{Z'}^2-m_l^2-2m_\pi E_{Z'}}{2(m_\pi-E_{Z'}+\sqrt{E_{Z'}^2-m_{Z'}^2})},~~~~~~~~~ E_e^{max}=\frac{m_\pi^2+m_{Z'}^2-m_l^2-2m_\pi E_{Z'}}{2(m_\pi-E_{Z'}-\sqrt{E_{Z'}^2-m_{Z'}^2})}.$$
The matrix element for the new decay mode of the pion is
\begin{equation}\label{matrix element}
{\cal M}=\frac{ f_{\pi } g_{\alpha\beta} G_F \cos \left(\theta _C\right)}{m_{\pi }^2-2 m_{\pi } E_e} p_\rho \bar{u}(k) \gamma^\mu(k\!\!\!/+q\!\!\!/)\gamma^\rho P_L v(l) \epsilon_\mu(q),
\end{equation}
where $G_F$, $\theta_C$,  and  $f_\pi$ are the Fermi constant, the Cabibbo angle, and the pion decay constant, respectively. $p$, $l$, $k$, and $q$ are the momentum of the pion, lepton, neutrino, and $Z'$, respectively.  $\epsilon_\mu$ is the polarization vector of $Z'$. Neglecting the mass of the final charged lepton leads to an error of $O(m_l^2/m_\pi^2)$ in the decay rate.  In the case of pion decay into an electron or a positron, the correction is less than $O(10^{-5})$ and is completely negligible.
With a summation over spins of the neutrino and electron and considering kinematics, we can write
\begin{equation}\label{Msquared1}
\sum_{spins}\vert{\cal M} \vert^2= f^2_{\pi } g^2_{\alpha\beta} G_F^2 \cos^2 \left(\theta _C\right)  tr(l\!\!\!/ \gamma^\mu k\!\!\!/ \gamma^\beta  P_L) \epsilon_\mu(q)   \epsilon_\beta^*(q).
\end{equation}

After a summation over the polarizations of $Z'$, we can write
\be \label{Msquared}\sum_{spins}\vert {\cal M} \vert^2=   g_{\alpha \beta}^2G_F^2f_\pi^2V_{qq'}^2\left(m_\pi^2+m_{Z'}^2-2m_\pi E_{Z'}+\frac{(m_\pi^2-m_{Z'}^2-2m_\pi E_l)(m_\pi^2-m_{Z'}^2-2m_\pi E_\nu)}{m_{Z'}^2}\right).
\ee
The total decay rate of pion [$\Gamma(\pi\longrightarrow e\nu_\alpha Z')$] is
\be
\Gamma(\pi\longrightarrow e\nu_\alpha Z')=\frac{  g_{e\alpha}^2 G_F^2 \cos^2 \left(\theta _C\right) f_\pi^2}{6144 \pi ^3 m_\pi^3 m_{Z'}^2}\left(m_\pi^8+72 m_\pi^4 m_{Z'}^4-64 m_\pi^2 m_{Z'}^6+24 \left(3 m_\pi^4 m_{Z'}^4+4 m_\pi^2 m_{Z'}^6\right) \log \left(\frac{m_{Z'}}{m_\pi}\right)-9
   m_{Z'}^8\right).
\ee
In the rest frame of the pion, the differential decay rate of the neutrino from $\pi\longrightarrow e\nu Z'$ to the energy of the neutrino is given by
\begin{equation}
\frac{d\Gamma(\pi\longrightarrow e\nu_\alpha Z')}{dE_\nu}=\frac{f_{\pi }^2 g_{e\alpha}^2 G_F^2 \cos^2 \left(\theta _C\right)}{64 \pi ^3 m_\pi^2  m_{Z'}^2 (m_\pi-2 E_\nu)^2}E_\nu^2 \left(2 E_\nu m_\pi-m_\pi^2+m_{Z'}^2\right)^2 \left(-2 E_\nu m_\pi+m_\pi^2+2 m_{Z'}^2\right).
\end{equation}
In the rest frame of the pion, the differential decay rates of the pion to the electron, neutrino and $Z'$ with  polarization  perpendicular to the $Z'$ momentum $(\epsilon_1,\epsilon_2)$ and parallel $(\epsilon_3)$
to the $Z'$ momentum are, respectively,\be
\frac{d\Gamma(\pi\longrightarrow e\nu_\alpha Z')}{dE_{Z'}} \mid_{1,2}=\frac{ f_{\pi }^2 g_{e\alpha}^2 G_F^2 \cos^2 \left(\theta _C\right)}{96 \pi ^3 m_\pi}p_{Z'} \left(-2 E_{Z'} m_\pi+m_\pi^2+m_{Z'}^2\right),
\ee
and
\be
\frac{d\Gamma(\pi\longrightarrow e\nu_\alpha Z')}{dE_{Z'}} \mid_{3}=\frac{f_{\pi }^2 g_{e\alpha}^2 G_F^2 \cos^2 \left(\theta _C\right)}{96 \pi ^3 m_\pi  m_{Z'}^2}p_{Z'} \left(E_{Z'} m_\pi-m_{Z'}^2\right)^2.
\ee
$Z'$ decays to a neutrino-antineutrino pair with the amplitude
\be
{\cal M}(Z'\longrightarrow \nu_\alpha\bar{\nu}_\beta) =g_{\alpha\beta} \bar{u}(q_1)\gamma^\mu P_L v(q_2) \epsilon_\mu(q),
\ee
in which, $q$, $q_1$ and $q_2$ are momenta of $Z'$, the neutrino and antineutrino respectively. With a summation over spins of neutrinos, the square of the amplitude is
\be
\sum_{spins}\vert{\cal M} \vert^2 =g^2_{\alpha\beta} tr(q_2\!\!\!/\gamma^\mu  q_1\!\!\!/   \gamma^\nu P_L )  \epsilon^*_\mu(q) \epsilon_\nu(q).
\ee
In the rest frame of $Z'$, we can write $q=(m_{Z'},0,0,0)$, $q_1=(E_\nu,E_\nu \sin\theta \cos\phi ,E_\nu \sin\theta \sin\phi,E_\nu \cos\theta)$ and $q_2=(E_\nu,-E_\nu \sin\theta \cos\phi,$ $-E_\nu \sin\theta \sin\phi,-E_\nu \cos\theta)$. In the case of $\epsilon_1$ polarization,
$$
\vert{\cal M} \vert^2=  4 g^2_{\alpha\beta} E_\nu^2(1-\sin^2\theta \cos^2\phi),
$$
in the case of $\epsilon_2$ polarization,
$$
\vert{\cal M} \vert^2=  4 g^2_{\alpha\beta} E_\nu^2(1-\sin^2\theta \sin^2\phi),
$$
and in the case of $\epsilon_3$,
$$
\vert{\cal M} \vert^2=  4 g^2_{\alpha\beta} E_\nu^2 \sin^2\theta.
$$
The partial decay rate is
$$
d\Gamma=\frac{1}{32\pi^2} \vert{\cal M} \vert^2 \frac{\vert k_1 \vert}{m^2_{Z'}} d\Omega.
$$
The total decay rate of the $Z'\longrightarrow\nu_\alpha\bar{\nu}_\beta$ for all the polarizations is equal to
$$
\Gamma(Z'\longrightarrow\nu_\alpha\bar{\nu}_\beta)=\frac{g^2_{\alpha\beta} m_{Z'}}{24\pi}.
$$
The neutrino spectrum produced from a polarized $Z^\prime$ decay in the rest frame of $Z'$ is given by
\be
(\frac{dN_\nu}{d\Omega})_{r.o.Z'}= \frac{1}{\Gamma(Z'\longrightarrow \nu \bar{\nu})} \frac{d\Gamma(Z'\longrightarrow \nu \bar{\nu})}{d \Omega},
\ee
In which in the case of $\epsilon_1$ and $\epsilon_2$, the normalized spectrum is
\be
(\frac{dN_\nu}{d\cos\theta})_{r.o.Z'}\mid_{1,2}=\frac{3(1+\cos^2\theta) }{8},
\ee
and in the case of $\epsilon_3$ is
\be
(\frac{dN_\nu}{d\cos\theta})_{r.o.Z'}\mid_{3}=\frac{3\sin^2\theta}{4}.
\ee
In the rest frame of the pion, the spectrum of neutrino produced from $Z'$ decay is
\be
(\frac{dN_\nu}{dE_\nu})_{r.o.\pi}\mid_{i}=(\frac{dN_\nu}{d\cos\theta})_{r.o.Z'}\mid_{i} \frac{d\cos\theta_{r.o.Z'}}{dE^\nu_{r.o.\pi}}= (\frac{dN_\nu}{d\cos\theta})_{r.o.Z'}\mid_{i} \frac{2}{E_{Z'} v_{Z'}},
\ee
in which  $v_{Z'}$ is the velocity of the $Z'$ in the rest frame of the pion.
For the $\epsilon_1$ and $\epsilon_2$ cases and for $Z'$ with a specific energy $E_{Z'}$,
\be
(\frac{dN_\nu}{dE_\nu})_{r.o.\pi}\mid_{1,2}= \frac{3 \left(-2 E_\nu^2 \left(4 E_\nu (E_\nu-E_{Z'})+m_{Z'}^2\right)-E_{Z'}^2 m_{Z'}^2+m_{Z'}^4\right)}{2 E_{Z'} \sqrt{1-\frac{m_{Z'}^2}{E_{Z'}^2}}
   \left(m_{Z'}^4-E_{Z'}^2 m_{Z'}^2\right)}
\ee
and for the $\epsilon_3$ polarization
$$
(\frac{dN_\nu}{dE_\nu})_{r.o.\pi}\mid_{3}=\frac{6 E_\nu^2 E_{Z'} \sqrt{1-\frac{m_{Z'}^2}{E_{Z'}^2}} \left(4 E_\nu (E_{Z'}-E_\nu)-m_{Z'}^2\right)}{\left(m_{Z'}^3-E_{Z'}^2 m_{Z'}\right)^2}.
$$
The total neutrino spectrum from $Z'$ decay in the rest frame of pion is given by
\be
(\frac{dN_\nu}{dE_\nu })_{r.o.\pi}^{Z'~ decay}= \sum_i\int_{E_\nu}^{E_{Z'}^{max}} dE_{Z'}    \frac{dN_{Z'}}{dE_{Z'}}\mid_{i} (\frac{dN_\nu}{dE_\nu})_{r.o.\pi}\mid_{i}
\ee
which $i$ refers to the $Z'$ polarization and
\be
\frac{dN_{Z'}}{dE_{Z'}}\mid_{i}=\frac{1}{\Gamma^\pi_{total}}\frac{d\Gamma(\pi\longrightarrow e\nu Z')}{dE_{Z'}}\mid_{i}.
\ee
With considering $\Gamma^\pi_{total}=\Gamma(\pi\longrightarrow\mu\nu_\mu)$, the total neutrino spectrum from $Z'$ decay in the rest frame of the pion is given by
$$(\frac{dN_\nu}{dE_\nu })_{r.o.\pi}^{Z' decay}=\frac{m_\pi \left(-2 E_\nu m_\pi+m_\pi^2+m_{Z'}^2\right) \left(4 E_\nu^3 m_\pi^2 (m_\pi-2 E_\nu)+4 E_\nu^2 m_\pi m_{Z'}^2
   (E_\nu-m_\pi)+m_\pi m_{Z'}^4 (m_\pi-2 E_\nu)+m_{Z'}^6\right)}{8 \pi ^2 m_{Z'}^4 m_\mu^2 \left(m_\mu^2-m_\pi^2\right)^2}.$$

The total spectrum of the electron (anti)neutrino produced from both pion and $Z'$ decay is given by

\begin{equation}\label{Eq.anuspec}
(\frac{dN_\nu}{dE_\nu})_{r.o.\pi}= (\frac{dN_\nu}{dE_\nu })_{r.o.\pi}^{Z' decay}
+ \frac{N_0}{ \Gamma (\pi\longrightarrow e\nu Z')}\frac{d\Gamma ( \pi\longrightarrow e\nu Z')}{dE_{\nu}}
\end{equation}
in which $N_0$ is the total number of the neutrinos produced from $\pi^+$ decay and it is equal to zero in the case of $\bar{ \nu} _e $ and non-zero  in the case of $\nu _e$ in $\pi^+$ decay mode.

In the lab frame the total spectrum of neutrino is given by

\be
(\frac{dN_\nu}{dE_\nu})_{lab}=(\frac{dN_\nu}{dE_{r.o.\pi}})_{r.o.\pi} \frac{ \partial E_{r.o.\pi}}{ \partial E_{lab}},
\ee
and in the case of the on-axis beam
$$
\frac{\partial E_{r.o.\pi}}{ \partial E_{lab}}=\gamma (1-v_\pi),$$
in which $v_\pi$ is the pion velocity in the lab frame.

Since the pion has a continuum spectrum, we calculate the neutrino flux from
\begin{equation}\label{Eq.nuflux}
\phi(E_\nu)= \frac{1}{4 \pi L^2} \int_{E_\pi^{min}}^{E_\pi^{max}}dE_{\pi} P_\pi(E_\pi) (\frac{dN_\nu}{dE_\nu})_{lab} \frac{d \Omega_{r.\pi}}{d \Omega_{lab}},
\end{equation}
where  $L$ is the distance from the source to the  detector and $P_\pi(E_\pi)$ is the rate of the pion  injection in the lab frame.
  $(\frac{dN_\nu}{dE_\nu})_{lab}$ is the spectrum of the neutrino in the lab frame from the decay of a pion with an energy of $E_\pi$, and $d \Omega_{r.\pi}/d\Omega_{lab}=(1+v_\pi)/(4(1-v_\pi)) \simeq \gamma_\pi^2$ takes care of focusing of the beam in the direction of the detector.


\end{document}